\documentclass[twocolumn,showpacs,preprintnumbers,superscriptaddress]{revtex4}

\usepackage{graphicx,dcolumn,bm,amsmath,amssymb}

\newcommand{\nl}{\nonumber \\}
\newcommand{\be}{\begin{equation}}
\newcommand{\ee}{\end{equation}}
\newcommand{\bea}{\begin{eqnarray}}
\newcommand{\eea}{\end{eqnarray}}
\newcommand{\bsube}{\begin{subequations}}
\newcommand{\esube}{\end{subequations}}
\newcommand{\Fig}[1]{Fig.\,\ref{#1}}
\newcommand{\Eq}[1]{Eq.\,(\ref{#1})}

\newcommand{\Sec}[1]{Section\,\ref{#1}}

\newcommand{\rmB}{{\rm B}}
\newcommand{\rmL}{{\rm L}}
\newcommand{\rmR}{{\rm R}}

\newcommand{\rmc}{{\rm c}}
\newcommand{\rmi}{{\rm i}}
\newcommand{\rmd}{{\rm d}}

\newcommand{\alf}{\alpha}
\newcommand{\sgm}{\sigma}
\newcommand{\Omg}{\Omega}
\newcommand{\omg}{\omega}
\newcommand{\Gam}{\Gamma}
\newcommand{\Dlt}{\Delta}
\newcommand{\dlt}{\delta}
\newcommand{\vpl}{\varepsilon}
\newcommand{\epl}{\epsilon}
\newcommand{\upa}{\uparrow}
\newcommand{\dwa}{\downarrow}
\newcommand{\GamL}{\Gamma_{\rm L}}
\newcommand{\GamR}{\Gamma_{\rm R}}

\newcommand{\Gamd}{\Gamma_{\rm d}}
\newcommand{\omgc}{\omega_{\rm c}}
\newcommand{\la}{\langle}
\newcommand{\ra}{\rangle}

\begin{document}

\title{Full counting statistics of renormalized dynamics in open quantum transport system}

 \author{JunYan Luo}
 \email{jyluo@zust.edu.cn}
 \affiliation{School of Science, Zhejiang University of Science
 and Technology, Hangzhou 310023, China}
 \author{Yu Shen}
 \affiliation{School of Science, Zhejiang University of Science
 and Technology, Hangzhou 310023, China}
 \author{Xiao--Ling He}
 \affiliation{School of Science, Zhejiang University of Science
 and Technology, Hangzhou 310023, China}
 \author{Xin--Qi Li}
 \affiliation{Department of Chemistry, Hong Kong University of Science and Technology, Kowloon,
 Hong Kong SAR, China}
 \affiliation{State Key Laboratory for Superlattices and Microstructures,
 Institute of Semiconductors, Chinese Academy of Sciences, P.O. Box
 912, Beijing 100083, China}
 \affiliation{Department of Physics, Beijing Normal University,
 Beijing 100875, China}
 \author{YiJing Yan}
 \affiliation{Department of Chemistry, Hong Kong University of Science and Technology, Kowloon,
 Hong Kong SAR, China}

\date{\today}

 \begin{abstract}
 The internal dynamics of a double quantum dot system is renormalized
 due to coupling respectively with transport electrodes and a
 dissipative heat bath. Their essential differences are identified
 unambiguously in the context of full counting statistics.
 The electrode coupling caused level detuning renormalization gives
 rise to a fast--to--slow transport mechanism, which is not resolved
 at all in the average current, but revealed uniquely by pronounced
 super--Poissonian shot noise and skewness.
 The heat bath coupling introduces an interdot coupling renormalization,
 which results in asymmetric Fano factor and an intriguing change of
 line shape in the skewness.
 \end{abstract}


\maketitle

 \section{\label{thsec1}Introduction}
 Electronic dynamics of an open quantum system has vital roles to play
 in the development of nanoelectronics \cite{Dat95,Lu03422,Fev071169}.
 Tremendous effort has been invested to characterize and control
 coherent dynamics in a variety of nanoscale transport
 systems \cite{Ono021313,Hay03226804,Pet052180,Kop06766,Fuj061634,Gus082547}.
 A lot of these studies are devoted to the energy relaxation and decoherence
 resulted from the coupling with a noisy
 environment \cite{Leg871,Wei08,Yan05187}.
 Yet, another important consequence of system--environment coupling, i.e.,
 internal energy renormalization, is attracting rising attention
 recently, owing to its essential
 influence on the internal dynamics of various nanoscale systems, such as spin
 valves \cite{Kon03166602,Bra06075328,Wun05205319,Sot10245319}, quantum
 dot (QD) Aharonov--Bohm interferometers \cite{Mar03195305,Bru96114},
 double quantum dots \cite{Fra04085301,Fra04L85},
 and the quantum state under measurement \cite{Luo09385801,Luo104904}.

 Full counting statistics (FCS), which characterizes the correlations
 between charge transport events of all orders \cite{Lev964845,Bag03085316},
 serves as a `spectroscopic' tool to reveal various internal
 mechanisms.
 In particular, owing to the development of highly sensitive on--chip
 detection of individual--electron tunneling technique, all statistical
 cumulants of the number of transferred particles can now be
 extracted experimentally.
 Electron transfer statistics in various nanoscale systems, such as
 QD \cite{Byl05361,Sch042005,Van044394,Gus06076605,Gus09191},
 gold atoms \cite{Gru073376}, and molecules \cite{Thi05045341,Wel081137},
 has been investigated and utilized to characterize the electronic
 dynamics.
 This offers new opportunities to study many--body effects such
 as the Coulomb blockade and Kondo dynamics with essential
 implications to nanoelectronics.

 The main purpose of the present work is to relate the higher
 cumulants to the renormalized dynamical properties that cannot
 be accessed by the average current measurements.
 To have a specific example, we will investigate the energy
 renormalizations in electron transport through a double quantum
 dot system \cite{Wie031}, where QD1 is directly 
 tunnel--coupled to the electrodes while QD2 is side connected to QD1,
 as schematically shown in \Fig{Fig1}.
 The double dots are further coupled to an inevitable dissipative
 phonon environment (not shown explicitly).
 The system is of particular interest, as it is in such a
 configuration that maximizes locality versus nonlocality
 contrast, and can be mapped onto nanostrucures in
 experiments \cite{Nau02161303,Sas09121926}.

 In context of FCS, we are capable of identifying the
 essential difference between the internal energy
 renormalizations induced by the coupling
 to the electrodes and that due to external phonon bath.
 The former one gives rise to a renormalized level detuning,
 which enhances both Fano factor and skewness uniquely  to a
 super--Poissonian value, whereas does not affect the average
 current at all.
 For the latter, i.e., the phonon bath coupling, it introduces
 an intriguing interdot coupling renormalization, causing strong
 electron localization in QD2.
 Furthermore, the asymmetry in phonon absorption and emission
 leads to an asymmetric shot noise spectrum as well as a change
 of line shape in the skewness.

 The remainder of this Letter is organized as follows.
 Based on the number--resolved reduced density matrix, the theory
 of FCS is outlined in \Sec{thsec2}.
 We then introduce in \Sec{thsec3} the model Hamiltonian for the
 transport double dots coupled to a heat bath.
 The physical origin of the internal energy renormalizations is
 discussed in \Sec{thsec4}, along with its influence on the FCS
 analyzed.
 Finally, all the results are summarized in \Sec{thsec5}.

 \section{\label{thsec2}FCS Formalism}

 Full information about transport properties of a given system
 is contained in the probability distribution $P(N,t_\rmc)$ that
 $N$ electrons have transferred through the system during the
 counting time $t_\rmc$.
 The FCS yields all the cumulants $\la I^k\ra$ of the probability
 distribution, which are directly related to important properties
 of the junction.
 For instance, the first cumulant $\la I\ra$ gives the
 average current. The shot noise is related to the second
 cumulant $\la I^2\ra$ and is commonly represented by the
 Fano factor $F\equiv \la I^2\ra/\la I\ra$, with $F<1$
 indicating a sub--Poisson noise, $F=1$ a Poisson fluctuation,
 and $F>1$ a super--Poisson process.
 The third cumulant
 $\la I^3\ra$ measures the skewness of the distribution.

 The cumulants can be found by performing derivatives of the
 cumulant generating function (CGF)
 ${\cal F}(\chi)$ with respect to the counting field $\chi$,
 \be\label{cumulant}
 \la I^k\ra=-\frac{1}{t_\rmc}(-\rmi\partial_\chi)^k{\cal F}(\chi,t_\rmc)|_{\chi=0},
 \ee
 where the CGF is defined as
 \be
 e^{-{\cal F}(\chi,t_\rmc)}\equiv\sum_{N}P(N,t_\rmc)e^{\rmi N\chi}.
 \ee
 The involving probability distribution is closely related to the dynamics
 of the reduced system $\rho(t)$, which is obtained from the
 density matrix of the entire system by integrating out the
 reservoir degrees of freedom. To characterize the FCS, we further
 unravel $\rho(t)$ into components $\rho^{(N)}(t)$, in which
 `$N$' denotes the number of electrons transferred through the system.
 The probability distribution is obtained via
 $P(N,t)={\rm Tr}\rho^{(N)}(t)$, where ${\rm Tr}(\cdots)$
 denotes the trace over the reduced system degrees of freedom.

 To proceed, instead of using $\rho^{(N)}(t)$ directly, we
 shall employ its $\chi$--space counterpart, i.e.,
 $\varrho(\chi,t)\equiv\sum_N \rho^{(N)}(t)e^{\rmi N\chi}$.
 The corresponding quantum master equation reads
 formally \cite{Li05205304,Luo07085325,Luo08345215,Wan07125416}
 \be\label{CQME}
 \frac{\partial}{\partial t}\varrho(\chi,t)=-(\rmi {\cal L}
 +{\cal R}_\chi)\varrho(\chi,t)\equiv{\cal L}_\chi\varrho(\chi,t),
 \ee
 where ${\cal L}(\cdots)\equiv[H_{\rm sys},(\cdots)]$ is the reduced system
 Liouvillian, ${\cal R}_\chi$ is the dissipation superoperator, and
 ${\cal L}_\chi\equiv-(\rmi {\cal L} +{\cal R}_\chi)$, which will be
 specified later in the localized state representation [see \Eq{QME}].

 With the knowledge of $\varrho(\chi,t)$, the CGF is determined
 straightforwardly as
 ${\cal F}(\chi,t_\rmc)=-\ln\{{\rm Tr}\varrho(\chi,t_\rmc)\}$.
 In the zero-frequency limit, i.e.\ the counting time $t_\rmc$ is
 much longer than the time of tunneling through the system,
 the CGF is simplified to \cite{Gro06125315,Fli05475,Kie06033312}
 \be
 {\cal F}(\chi,t_\rmc)=-\lambda_{\rm min}(\chi) t_\rmc,
 \ee
 where $\lambda_{\rm min}(\chi)$ is the minimal eigenvalue
 of ${\cal L}_\chi$ that satisfies
 $\lambda_{\rm min}(\chi\rightarrow0)\rightarrow0$.

 \section{\label{thsec3}Model description}

 \begin{figure}
 \begin{center}
 \includegraphics*[scale=0.65]{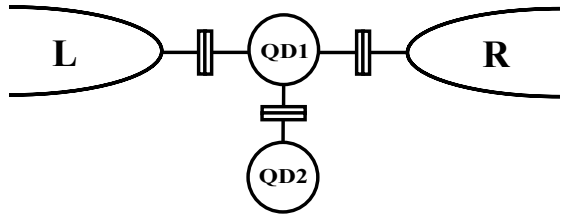}
 \caption{\label{Fig1}Schematic setup for transport through a
 double quantum dot system, where QD1 is tunnel--coupled to the
 left and right electrodes, while QD2 is side--connected to the QD1.
 The system is further coupled to an inevitable dissipative phonon
 environment (not shown explicitly).}
 \end{center}
 \end{figure}

 Specifically, let us consider a double quantum dot system,
 as schematically shown in \Fig{Fig1}, where the QD1 is connected
 to left and right electrodes, while the QD2 is side--connected
 to the QD1. The double dots are furthermore embedded in a
 dissipative phonon bath. The total Hamiltonian consists of the
 double dot system, the environment, and the coupling between them,
 \be\label{Htot}
 H=H_{\rm sys}+H_{\rm env}+H_{\rm sys-env}.
 \ee
 Here the first part models the coupled dots
 \bea\label{Hs}
 H_{\rm sys}=&\!\!\!\sum_{\sgm}\left[\left(E_0+\frac{1}{2}\epl
 \hat{Q}_{z\sgm}\right)+\Omg \hat{Q}_{x\sgm}\right]
 \nl
 &\!\!\!+\sum_{\ell=1,2}U_0\hat{n}_{\ell\upa}\hat{n}_{\ell\dwa}
 +U'\hat{n}_1 \hat{n}_2,
 \eea
 where $\hat{Q}_{z\sgm}\equiv d_{1\sgm}^\dag d_{1\sgm}-d_{2\sgm}^\dag d_{2\sgm}$,
 $\hat{Q}_{x\sgm}\equiv d_{1\sgm}^\dag d_{2\sgm}+d_{2\sgm}^\dag d_{1\sgm}$,
 $\hat{n}_\ell=\sum_\sgm \hat{n}_{\ell\sgm}$, and
 $\hat{n}_{\ell\sgm}=d_{\ell\sgm}^\dag d_{\ell\sgm}$,
 with $d_{\ell\sgm}$ ($d_{\ell\sgm}^\dag$) the electron
 annihilation (creation) operator in the QD1 ($\ell$=1)
 or QD2 ($\ell$=2) and spin $\sgm=\upa$ or $\dwa$.
 Each dot consists of a single spin--degenerate energy level
 $E_{1/2}=E_0\pm\frac{1}{2}\epl$, measured relative to the
 equilibrium chemical potential of the electrodes. Electron
 tunneling between the two dots are characterized by the
 interdot tunneling strength $\Omg$. Double occupation on
 the same dot costs the intradot charging energy $U_0$.
 Simultaneous occupation of one electron in each dot is
 associated with the interdot charge energy $U'$. Hereafter,
 the intradot charging energy $U_0$ is assumed to be much
 larger than the bias voltage, such that charge states with
 three or more electrons in the double dot are prohibited.

 The environment is composed of the phonon bath, the left and
 right electrodes. The corresponding Hamiltonian reads
 $H_{\rm env}=h_{\rm ph}+h_{\rm el}$.
 Each of them is modeled as a collection of noninteracting
 particles.
 The phonon bath are modeled with
 $h_{\rm ph}=(1/2)\sum_j\hbar\omg_j (p_j^2+x_j^2)$.
 The electrodes assumes
 $h_{\rm el}=\sum_{\alf=\rmL,\rmR}
 \sum_{k\sgm}\vpl_{\alf k}c^\dag_{\alf k\sgm}c_{\alf k\sgm}$,
 expressed in terms of electron creation $(c^\dag_{\alf k\sgm})$
 and annihilation ($c_{\alf k\sgm}$) operators in the left
 ($\alf=$L) or right ($\alf=$R) electrode. The electrodes are
 assumed to be in equilibrium, so that they are characterized
 by the Fermi distribution  $f_{\rm L/R}(\omg)$.
 An applied bias voltage $V$ is modeled by different chemical
 potentials in the left and right electrodes $\mu_{\rm L/R}=\pm eV/2$.

 The system--environment coupling can be written as
 $H_{\rm sys-env}=H_{\rm sys-el}+H_{\rm sys-ph}$, with
 \bsube
 \bea
 H_{\rm sys-el}\!\!&=&\!\!\sum_{\alf k\sgm}(t_{\alf k}c_{\alf k\sgm}^\dag d_{1\sgm}
 +{\rm H.c.}),\label{Hsysel}
 \\
 H_{\rm sys-ph}\!\!&=&\!\!\sum_{\sgm}\hat{Q}_{z\sgm} F_{\rm ph}.
 \eea
 \esube
 Here $H_{\rm sys-el}$ describes electron transfer between QD1 and
 the electrodes.
 $H_{\rm sys-ph}$ models the coupling with phonon bath, in
 which $F_{\rm ph}=\sum_j \lambda_jx_j$. This term is
 responsible for the dot level energy fluctuations.
 The tunnel coupling between electrodes and QD1 is
 characterized by the coupling strength
 $\Gam_\alf=2\pi\sum_k|t_{\alf k}|^2\dlt(\omg-\vpl_{\alf k}) $.
 In what follows, we consider only spin conserving tunneling
 processes, and assume wide bands in the reservoirs, which
 yields energy independent couplings $\Gam_\alf$.
 The effect of phonon bath on the double dot system is
 characterized by the phonon interaction spectral density,
 $J_{\rm ph}(\omg)=\sum_j |\lambda_j|^2\dlt(\omg-\omg_j)$.
 Hereafter, it is assumed to be Ohmic, i.e.,
 $J_{\rm ph}(\omg)=\eta \omg e^{-\omg/\omgc}$, where the
 dimensionless parameter $\eta$ reflects the strength of
 dissipation and $\omgc$ is the Ohmic high energy cutoff.
 Throughout this work, we set unit of $\hbar=e=1$ for the
 Planck constant and electron charge, unless stated otherwise.

 \section{\label{thsec4}Level renormalization and FCS analysis}

 In the strong intradot Coulomb blockade regime, double occupation
 on the same dot is prohibited. The involving states are restricted
 to: $|0\ra$--both dots empty, $|1\sgm\ra$--one electron in QD1,
 $|2\sgm\ra$--one electron in QD2, and $|1\sgm2\sgm'\ra$--one
 electron in each dot, respectively.
 The quantum master equation (\ref{CQME}) in this localized state
 representation reads
 \bsube\label{QME}
 \begin{align}
 &\dot{\varrho}_{0}=-2(\Gam_{\rmL}^{+}+\Gam_{\rmR}^{+})\varrho_{0}+
 (\Gam_{\rmL}^{-}+\Gam_{\rmR}^{-}e^{\rmi\chi})(\varrho_{1\upa}+\varrho_{1\dwa}),
 \\
 &\dot{\varrho}_{1\sgm}=\rmi \Omg (\varrho^{1\sgm}_{2\sgm}-\varrho^{2\sgm}_{1\sgm})
 -(\GamL^-+\GamR^-)\varrho_{1\sgm}
 \nl &\qquad+(\GamL^++\GamR^+e^{-\rmi\chi})\varrho_0,
 \\
 &\dot{\varrho}_{2\sgm}=\rmi\Omg (\varrho^{2\sgm}_{1\sgm}-\varrho^{1\sgm}_{2\sgm})
 -2(\tilde{\Gam}_{\rmL}^++\tilde{\Gam}_{\rmR}^+)\varrho_{2\sgm}
 \nl &\qquad+(\tilde{\Gam}_{\rmL}^-\!+\tilde{\Gam}_{\rmR}^-e^{\rmi\chi})(\varrho_{1\sgm2\sgm}
 \!+\!\varrho_{1\bar{\sgm}2\sgm}),
 \\
 &\dot{\varrho}_{1\sgm2\sgm'}
 =-(\tilde{\Gam}_{\rmL}^-+\tilde{\Gam}_{\rmR}^-)\varrho_{1\sgm2\sgm'}
 +(\tilde{\Gam}_{\rmL}^++\tilde{\Gam}_{\rmR}^+e^{-\rmi\chi})\varrho_{2\sgm'},
 \\
 & \dot{\varrho}^{1\sgm}_{2\sgm}=\rmi(\epl+\tilde{\epl})\varrho^{1\sgm}_{2\sgm}
 +\rmi(\Omg+\tilde{\Omg})(\varrho_{1\sgm}-\varrho_{2\sgm})
 \nl &\qquad+\Gamma_+\varrho_{1\sgm}
 -\Gamma_-\varrho_{2\sgm}-\Gamd\varrho^{1\sgm}_{2\sgm},\label{off-diag}
 \end{align}
 \esube
 with spin $\sgm\in\{\upa,\dwa\}$ and $\bar{\sgm}=-\sgm$.
 Here $\varrho_s=\la s|\varrho|s\ra$ represents the diagonal element
 of the reduced density matrix. The off--diagonal elements
 $\varrho^s_{s'}=\la s|\varrho|s'\ra$ describes the so--called
 quantum ``coherencies''.
 The involving temperature--dependent tunneling rates are defined
 as $\Gam_\alf^{\pm}=\Gam_\alf f_\alf^{(\pm)}(E_0)$ and
 $\tilde{\Gam}_\alf^{\pm}=\Gam_\alf f_\alf^{(\pm)}(E_0+U')$,
 where $f^{(\pm)}_\alf(\omg)=\{1+e^{\pm \beta(\omg-\mu_\alf)}\}^{-1}$
 is related to the Fermi function of the electrode $\alf$= L or R,
 with $\beta=(k_\rmB T)^{-1}$.
 Here we are interested in the regime $\Dlt<k_\rmB T$
 ($\Dlt\equiv\sqrt{\epl^2+4\Omg^2}$ being the eigenenergy separation),
 where the external coupling strongly modifies the internal dynamics,
 and the off--diagonal elements of the reduced density matrix
 have vital roles to play \cite{Gur9615932,Sto961050}.
 The level separation is thus smeared by the temperature and only
 the excitation energy levels $E_0$ and $E_0+U'$ enter the Fermi
 functions.

 The coupling to the phonon bath results in relaxation and
 dephasing in the system, with corresponding rates
 given by \cite{Bra05315,Kie07206602,Agu04206601,Luo10083720}
 \bsube
 \be
 \Gam_{\pm}=\frac{\eta\pi\Omg\epl}{\beta\Dlt^2}
 -\frac{\pi}{2}\frac{\Omg}{\Dlt}J_{\rm ph}(\Dlt)
 \left[\frac{\epl}{\Dlt}\coth\left(\frac{\beta\Dlt}{2}\right)\pm1\right],
 \ee
 and
 \be
 \gamma_{\rm ph}=2\pi\frac{\Omg^2}{\Dlt^2}J_{\rm ph}(\Dlt)
 \coth\left(\frac{\beta\Dlt}{2}\right)+\frac{\eta\pi\epl^2}{\beta\Dlt^2}.
 \ee
 \esube
 The total dephasing rate eventually reads
 \be
 \Gamd={\textstyle \frac{1}{2}}(\GamL^-+\GamR^-)+
 (\tilde{\Gam}_{\rmL}^++\tilde{\Gam}_\rmR^+)+\gamma_{\rm ph}.
 \ee

 By observing the equation of motion of the off--diagonal matrix
 element [\Eq{off-diag}], it is found that the energy scales of
 the double dots are renormalized, i.e.,
 $\epl\rightarrow\epl+\tilde{\epl}$ and
 $\Omg\rightarrow\Omg+\tilde{\Omg}$.
 The energy renormalizations $\tilde{\epl}$ and $\tilde{\Omg}$
 are entailed respectively by coupling with the electrodes and
 phonon bath,
 \bsube
 \bea
 \tilde{\epl}\!\!\!&=&\!\!\!\phi(E_0)-2\phi(E_0+U')+\phi(E_0+U_0),
 \\
 \tilde{\Omg}\!\!\!&=&\!\!\!2\pi\rmi\frac{\eta\Omg}{\beta \Dlt}\{D(\Dlt)-D(-\Dlt)\},
 \eea
 \esube
 with
 \bsube
 \bea
 \phi(\omg)\!\!\!&=&\!\!\!\sum_{\alf=\rmL,\rmR}\frac{\Gam_\alf}{2\pi}{\rm Re}
 \left[\Psi\left(\frac{1}{2}
 +\rmi\,\beta \frac{\omg-\mu_\alf}{2\pi}\right)\right],
 \\
 D(\omg)\!\!\!&=&\!\!\!e^{-\rmi\frac{\omg}{\omgc}}B_x(y,-1)
 -e^{\rmi\frac{\omg}{\omgc}}B_x(y^\ast,-1).
 \eea
 \esube
 Here, $\Psi$ and $B_x(y,z)$ are respectively digamma
 function and  incomplete beta function \cite{Gra80}, with
 $x=e^{-\frac{2\pi}{\beta\omgc}}$ and
 $y=1-\rmi\frac{\beta\omg}{2\pi}$.

 \begin{figure*}
 \begin{center}
 \includegraphics[scale=0.65]{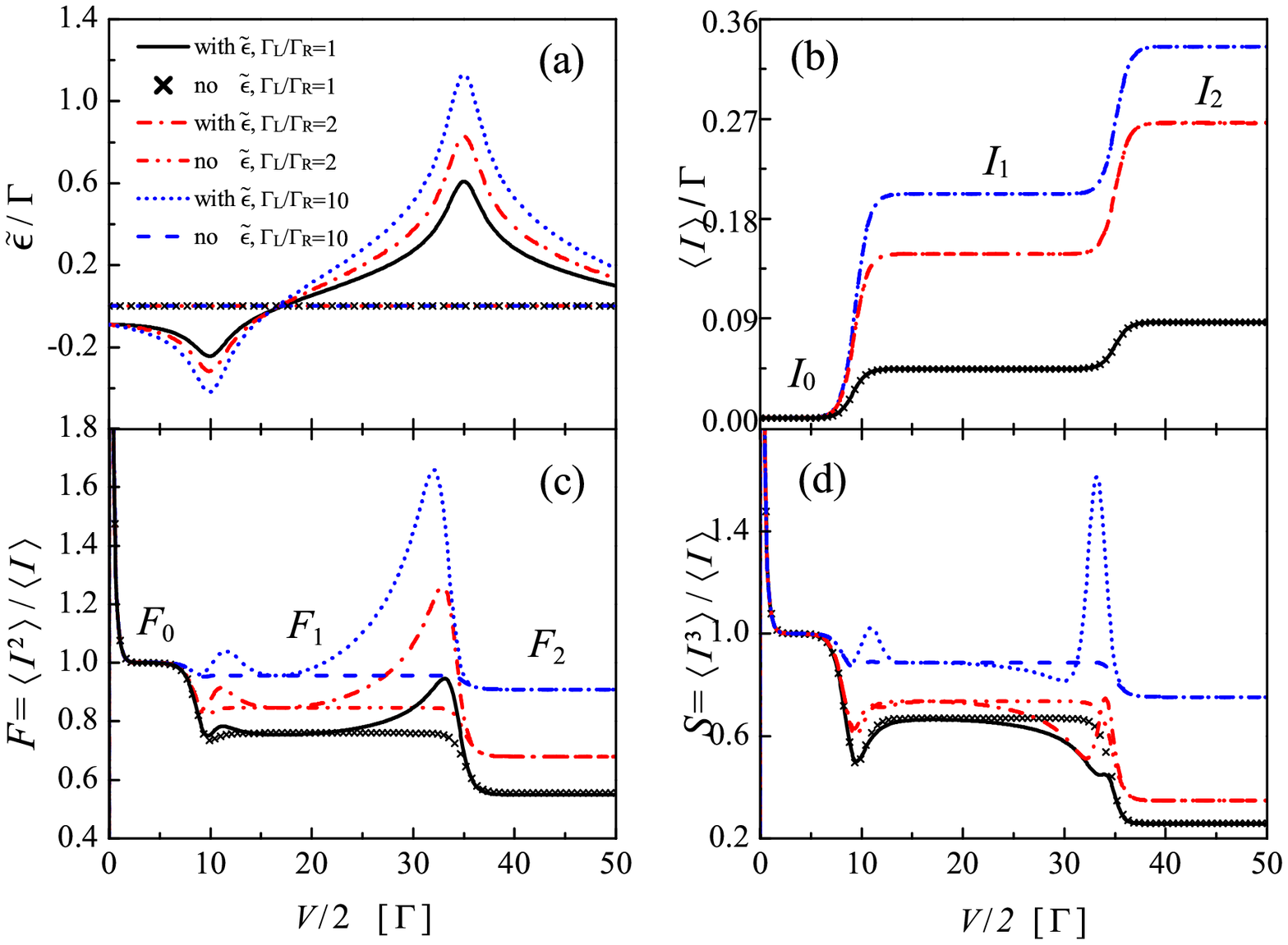}
 \caption{\label{Fig2} (a) level detuning renormalization, (b) average
 current $\la I\ra$, (c) Fano factor $F=\la I^2\ra/\la I\ra$, and (d)
 normalized skewness $S=\la I^3\ra/\la I\ra$  versus the bias voltage
 for different $\GamL/\GamR$ ratios.
 Each time when the chemical potential of the electrode aligns
 with the energy needed for either single ($E_0$) or double
 occupation ($E_0+U'$), $\tilde{\epl}$ reaches its local extremum.
 Plot parameters are: $\epl$=0, $\Gam\equiv\GamL+\GamR=2\Omg$,
 $E_0=10k_\rmB T$, $U'=25k_\rmB T$, and $U_0=60k_\rmB T$.}
 \end{center}
 \end{figure*}

 The detuning renormalization $\tilde{\epl}$ has been neglected in
 previous works \cite{Gur9615932,Gur9715215,%
 Gur986602,Sto961050,Haz01165313},
 where the Fermi energies of the electrodes are assumed to be far
 away from the electronic states of the dots.
 The effect of interdot coupling renormalization, to our knowledge,
 has not yet been addressed in the context of electron quantum
 transport.

 To clearly address the interdot coupling renormalization, we shall
 employ the quantum master equation (\ref{CQME}), in which the
 dissipative term contains the contribution from the phonon bath
 \be
 {\cal R}_{\rm ph}\varrho(\chi,t)
 =[\hat{Q}_{z},\tilde{Q}_{z}\varrho(\chi,t)]+{\rm H.c.},
 \ee
 with $\hat{Q}_{z}=\sum_\sgm\hat{Q}_{z\sgm}$ and
 \be\label{tiQ}
 \tilde{Q}_{z}=[\tilde{C}(-{\cal L})+\rmi\tilde{D}(-{\cal L})]\hat{Q}_{z}.
 \ee
 Here, $\tilde{C}(-{\cal L})=\int_{-\infty}^{\infty}\rmd t C(t) e^{-\rmi{\cal L}t}$
 is spectral function.
 The involving phonon bath correlation function is
 $C(t)=\la F_{\rm ph}(t)F_{\rm ph}(0)\ra_{\rm B}$, with
 $\la \cdots \ra\equiv{\rm Tr}_{\rm B}[(\cdots)\rho_{\rm B}]$, and
 $\rho_{\rm B}$ the thermal equilibrium state of the phonon bath.
 The dispersion function $\tilde{D}(-{\cal L})$ involved can be evaluated
 via the Kramers--Kronig relation
 $\tilde{D}(-{\cal L})=-\frac{1}{\pi}{\cal P}\int_{-\infty}^{\infty} \rmd \omg
 \frac{C(\omg)}{{\cal L}-\omg}$,
 where ${\cal P}$ denotes the principal value. This term is responsible
 for the energy renormalization \cite{Yan05187}.
 Let us focus on the
 last term in \Eq{tiQ}, which can be further expressed as
 \be
 \rmi\tilde{D}(-{\cal L})\hat{Q}_{z}=4\pi\rmi\frac{\eta\Omg^2}{\beta\Dlt^2}
 [D(\Dlt)+D(-\Dlt)]\hat{Q}_{z}+\tilde{\Omg}\hat{Q}_{+}-\tilde{\Omg}\hat{Q}_{-},
 \ee
 with $\hat{Q}_{+}=|1\ra\la2|$ and $\hat{Q}_{-}=|2\ra\la1|$.
 Apparently, the first term accounts for the renormalization of
 level detuning. However, due to the odd function $D(\omg)$,
 this term eventually vanishes.
 The last two terms represent phonon absorption and emission.
 It is this phonon-induced effective coupling that gives rise
 to a unique interdot coupling modification which takes place
 only in \Eq{off-diag}.

 Hereafter, we will show that the both $\tilde{\epl}$
 and $\tilde{\Omg}$ have important roles to play in the reduced
 dynamics, and consequently give rise to intriguing features in
 the high order cumulants of the current distribution.
 Consider first the situation without electron--phonon
 interaction ($\eta=0$). The renormalization of interdot
 coupling $\tilde{\Omg}$ is thus zero.
 The numerical result for $\tilde{\epl}$ vs bias voltage is
 shown in \Fig{Fig2}(a) for different $\GamL/\GamR$ ratios.
 Once the Fermi energy of the electrode is resonant with
 the energy needed for single ($E_0$) or double occupation
 ($E_0+U'$), $\tilde{\epl}$ reaches its local extremum.
 For a fixed total tunneling width $\Gam=\GamL+\GamR$,
 $\tilde{\epl}$ increases whenever the tunnel--coupling
 asymmetry $\GamL/\GamR$ increases.

 \begin{figure}
 \begin{center}
 \includegraphics*[scale=0.46]{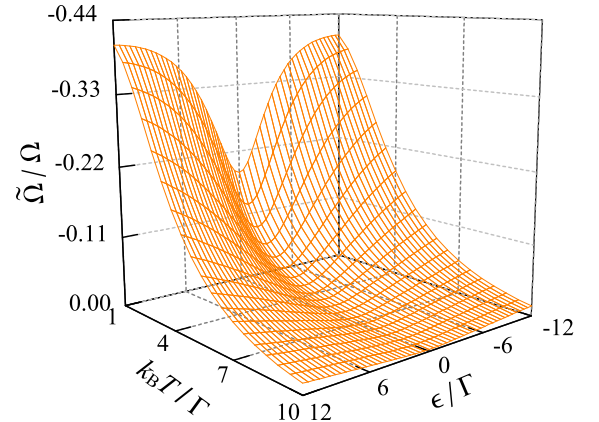}
 \caption{\label{Fig3} 3D plot of the interdot coupling renormalization
 $\tilde{\Omg}$ vs bare level detuning $\epl$ and temperature
 $k_\rmB T$.
 The plotting parameters are $\GamL=\GamR=\Gam/2$, and $\eta=0.04$,
 $\omgc=15\Gam$ for the Ohmic phonon bath spectral density.}
 \end{center}
 \end{figure}

 Let us now investigate the influence of level detuning
 renormalization on FCS.
 The first three cumulants of the transport current are
 displayed in \Fig{Fig2}(b), (c), and (d), respectively,
 as functions of the bias voltage.
 Here, the results with bare level detuning ($\tilde{\epl}=0$)
 are also plotted for comparison.

 The first cumulant, i.e., average current shows a typical step-like
 structure, as displayed in \Fig{Fig2}(b).
 In the low bias voltage regime ($\mu_\rmL,\mu_\rmR<E_0$), transport is
 exponentially suppressed, as the Fermi energies of both electrodes are
 well below the single level $E_0$. There are only very few thermally
 activated tunneling events, and the current $I_0\approx 0$.
 Transport through the system becomes energetically allowed when the
 Fermi level crosses the discrete level, i.e., $\mu_\rmL
 >E_0>\mu_\rmR$. The current rises to the second plateau.
 The bias voltage, however, is not sufficient to overcome the intradot
 and interdot charging energies, and the coupled dots can accommodate
 at most one electron. This is the so--called double--dot Coulomb
 blockade (DDCB) regime \cite{Luo08345215,Luo10083720}.
 The stationary current is obtained by utilizing \Eq{cumulant}
 \be\label{Iddcb}
 I_1=\frac{2\GamL\GamR}{4\GamL+\GamR},
 \ee
 where the involving Fermi functions are approximated by either one
 or zero. The effective quadrupling of $\GamL$ is a combined effect
 of spin degeneracy and DDCB \cite{Luo07085325}.

 As the bias further increase and cross interdot charging energy,
 i.e.,  $\mu_\rmL >E_0+U',E_0>\mu_\rmR$, double occupancy (with
 one electron in each dot) is possible. It is also referred
 to as single--dot Coulomb blockade (SDCB) regime.
 The current eventually rises to the third plateau as shown in
 \Fig{Fig2}(b), with its value given by
 \be
 I_2=\frac{2\GamL\GamR}{2\GamL+\GamR},
 \ee
 where the effective doubling of $\GamL$ arises from the spin
 degeneracy and SDCB \cite{Luo07085325}.

 Noticeably, the current is not sensitive at all to the level
 detuning renormalization $\tilde{\epl}$.
 The reason will be provided later.
 In order to further exploit the effect of the energy renormalization,
 we resort to higher order cumulants.
 The numerical result of the Fano factor against the bias voltage is
 displayed in \Fig{Fig2}(c).
 At very small bias $V\ll k_{\rm B}T$, it is dominated by thermal noise,
 which is described by the well--known hyperbolic cotangent behavior, and
 eventually leads to a divergence of the Fano factor at $V=0$
 \cite{Bla001,Suk01125315}.
 As the bias increases but still below $E_0$, there are only
 thermally activated uncorrelated tunneling events, and Fano
 factor is Poissonian, i.e., $F_0=1$.
 In the DDCB regime, electron transport through the bare level $E_0$ is
 energetically allowed.
 The Fano factor is given by
 \be
 F_1=1-\frac{8\GamL\GamR}{(4\GamL+\GamR)^2}
 +\frac{2\GamL^2(\Gam_\rmR^2+4\tilde{\epl}^2)}{(4\GamL+\GamR)^2\Omg^2},
 \ee
 where the second term describes the suppression of the Fano factor
 below unity, whereas the third term gives a positive contribution.
 Unambiguously, the Fano factor is modulated by the detuning
 renormalization $\tilde{\epl}$.
 This shows the shot noise as a much more sensitive tool to the
 system parameters than the average current.

 The level detuning renormalization gives rise to intriguing features
 in the noise spectrum.
 By neglecting $\tilde{\epl}$, the noise in the DDCB regime is constant
 against the bias. It can reach maximally the Poissonian value
 (for the chosen parameters here $\GamL+\GamR=2\Omg$) in the
 limit of large tunnel--coupling asymmetry, i.e., $\GamL/\GamR\gg
 1$.
 The detuning renormalization enhance the Fano factor in a unique
 way, i.e., the noise are increased at bias close to excitation
 energies $E_0$ and $E_0+U'$.
 Remarkably, a pronounced super--Poissonian noise is observed provided
 tunnel couplings are sufficient asymmetric, as shown by the dotted curve
 in \Fig{Fig2}(c).

 To further understand the current and noise features, we perform a
 unitary transformation of the entire Hamiltonian, such that the
 system Hamiltonian is diagonalized to (here spin indices are suppressed
 for simplicity)
 \be
 \tilde{H}_{\rm sys}=\frac{1}{2}\tilde{\Dlt}\{|+\ra\la+|-|-\ra\la-|\},
 \ee
 where $\tilde{\Dlt}\equiv\sqrt{\tilde{\epl}^2+4\Omg^2}$, with the
 detuning renormalization $\tilde{\epl}$ being appropriately taken into
 account.
 The coupled dot system can now be mapped onto a parallel two--level
 system. Here, the bonding state $|+\ra$ and the anti--bonding state
 $|-\ra$ are defined respectively as
 \bsube
 \bea
 |+\ra=\sin\frac{\theta}{2}|2\ra+\cos\frac{\theta}{2}|1\ra,
 \\
 |-\ra=\cos\frac{\theta}{2}|2\ra-\sin\frac{\theta}{2}|1\ra,
 \eea
 \esube
 where $\theta$ is introduced via $\sin\theta=2\Omg/\tilde{\Dlt}$
 and $\cos\theta=\tilde{\epl}/\tilde{\Dlt}$.
 As a result, the electron tunnel--coupling Hamiltonian (\ref{Hsysel})
 is recast to
 \be
 \tilde{H}_{\rm sys-el}=\sum_{\alf k}\{t_{\alf k}^+
  c_{\alf k}^\dag|0\ra\la+|
  +t_{\alf k}^-c_{\alf k}^\dag|0\ra\la-|\}+{\rm H.c.},
 \ee
 with $t_{\alf k}^+=t_{\alf k}\cos\frac{\theta}{2}$ and
 $t_{\alf k}^-=t_{\alf k}\sin\frac{\theta}{2}$.
 Apparently, the amplitudes for tunneling through the eigenstates
 $|+\ra$ and $|-\ra$ are effectively modulated by the detuning
 renormalization $\tilde{\epl}$.

 The stationary current in general can be written as
 \be
 I=\int \frac{d\omg}{2\pi} {\cal T}(\omg) \{f_\rmL(\omg)-f_\rmR(\omg)\},
 \ee
 where ${\cal T}(\omg)$, the transmission probability, defines the
 probability for an incoming electron to be transmitted through
 the device. For a parallel two--level system in the DDCB regime,
 the transmission probability is related to
 \bea\label{Tomg}
 {\cal T}&\sim & \frac{(|t_{\rmL k}^+|^2+|t_{\rmL k}^-|^2)|t_{\rmR k}^+t_{\rmR k}^-|^2}
 {|t_{\rmR k}^+t_{\rmR k}^-|^2+2|t_{\rmR k}^+t_{\rmL k}^-|^2+2|t_{\rmR k}^-t_{\rmL k}^+|^2}
 \nl
 &= & \frac{2|t_{\rmL k}|^2|t_{\rmR k}|^2}{4|t_{\rmL k}|^2+|t_{\rmR
 k}|^2},
 \eea
 where the $\theta$--dependence of the tunneling amplitudes cancel out.
 Eventually, it leads to the current expression of \Eq{Iddcb}.
 Physically, suppression of electron tunneling through the
 anti--bonding state due to level renormalization is exactly compensated
 by that through the bonding state.
 The total transmission probability thus remains unchanged, and finally
 the current turns out to be detuning renormalization independent.

 The Fano factor, however, is more sensitive to the internal
 dynamics, and reveals much detailed information about transport.
 Particularly, in the presence of large $\tilde{\epl}$,
 electron tunneling through the anti--bonding state is suppressed,
 while that through bonding state is enhanced.
 In the limit where the Coulomb interactions prevent a double
 occupancy of system, there is competition between the two
 transport channels. Consequently, the slow flowing of electrons
 through the bonding state modulates that through the fast state,
 giving rise to a bunching of tunneling events and
 eventually resulting in super--Poissonian noise, as displayed by the
 dotted curve in \Fig{Fig2}(c).

 Different from the Fano factor, the skewness can be either increased
 or reduced by $\tilde{\epl}$.
 In the case of symmetric tunnel--coupling ($\GamL=\GamR$), the level
 renormalization leads to a suppression of skewness, see the
 solid curve in \Fig{Fig2}(d).
 As the coupling asymmetry grows, $\tilde{\epl}$ can strongly enhance
 the skewness to a super--Poissonian value, and pronounced noise peaks
 are observed at bias close to $E_0$ and $E_0+U'$, as shown by the
 dotted curve in \Fig{Fig2}(d).
 The peak width of skewness, however, is reduced remarkably,
 compared with that of Fano factor.

 We now investigate the effect of energy renormalization induced by
 external phonon bath.
 Let us ficus on the DDCB regime, i.e., at most one electron can reside
 on the double dot system, and investigate the effect of $\tilde{\Omg}$
 based on FCS.
 One can always choose an appropriate bias voltage, at which the level
 detuning renormalization is zero, i.e., $\tilde{\epl}=0$, such that
 its influence is eliminated.
 In \Fig{Fig3}, $\tilde{\Omg}$ is plotted as a
 function of temperature $k_\rmB T$ and bare level detuning $\epl$.
 It is observed that the interdot coupling renormalization
 is fully negative; it thus reduces the total interdot coupling.
 It reaches a local minimum at $\epl=0$, and grows with increasing
 $\epl$.
 $\tilde{\Omg}$ is particularly notable at low temperature, but falls
 off rapidly with rising temperature.

\begin{figure}
 \begin{center}
 \includegraphics*[scale=0.5]{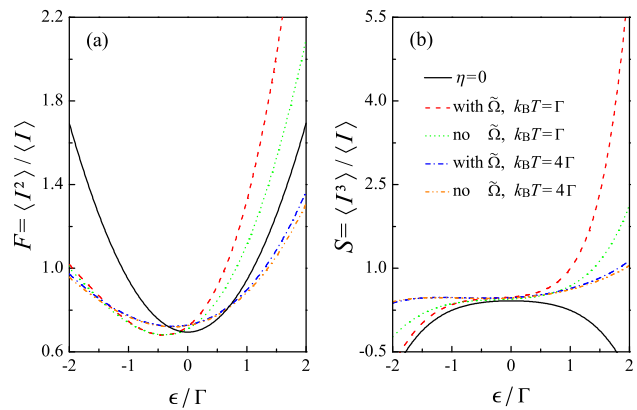}
 \caption{\label{Fig4} (a) Fano factor and (b) normalized skewness
 in the DDCB regime vs bare level detuning for different temperatures.
 The other plotting parameters are the same as those in \Fig{Fig3}.}
 \end{center}
 \end{figure}

 The calculated Fano factor and normalized skewness are shown in
 \Fig{Fig4}(a) and (b), respectively, as a function of the bare level
 detuning at different temperatures.
 For comparison, the results without $\tilde{\Omg}$ are also
 plotted.
 In the absence of the phonon bath ($\eta=0$), both Fano factor
 and skewness are symmetric around $\epl=0$ (see the solid
 curves in \Fig{Fig4}).
 With nonzero coupling to the heat bath, phonon absorption and
 emission take place, which dominate for $\epl<0$ and $\epl>0$,
 respectively.
 At low temperatures, the asymmetry in emission and absorption
 leads to the observed asymmetric spectra in \Fig{Fig4}.
 The coupling to the phonon bath furthermore causes dephasing
 between the two dot--levels, which generally suppresses the
 noise \cite{Kie07206602,Bra091048}. The reduction of Fano
 factor is particularly notable for $\epl<0$, as shown in
 \Fig{Fig4}(a).
 On the other hand, electron--phonon scattering results in electron
 localization in QD2 \cite{Luo10083720}.
 In particular, for $\epl>0$ and at low temperature (e.g. $k_\rmB T=\Gam$)
 this mechanism dominates. Electrons thus tend to be transferred in
 bunches, which enhances the noise in comparison with the case of
 $\eta=0$, as displayed by the dotted curves in \Fig{Fig4}(a).

 Finite $\tilde{\Omg}$ (due to its negative contribution) suppresses
 electron tunneling between the two dots, and effectively
 increases electron localization in QD2. Thereby the noise is enhanced,
 particularly at low temperature, where $\tilde{\Omg}$ is most
 prominent [see the dashed curves in \Fig{Fig4}(a)].
 Noise enhancement due to weak interdot coupling were also reported
 in Ref. \cite{Urb09165319}, where counting statistics for electron
 transport through a double-dot
 Aharonov--Bohm interferometer was investigated. However, there the
 divergence is closely related to a separation of the Hilbert space
 of the double dots into disconnected subspaces that contain the spin
 singlet and triplet states for double occupancy.

 The effect of phonon bath on the third order cumulant is displayed
 in \Fig{Fig4}(b), where the normalized skewness is plotted against
 the bare detuning $\epl$.
 Whenever the double dots are coupled to the phonon bath, the
 skewness is enhanced and becomes asymmetric.
 Noticeably, above the resonance ($\epl>0$) a change of line shape
 is observed, i.e., the skewness increases with growing $\epl$,
 rather than decreases as in the case of $\eta=0$.
 This intriguing feature shows unambiguously the third order cumulant
 as a much sensitive tool to the phonon bath coupling.
 The presence of finite $\tilde{\Omg}$ furthermore
 increases the skewness, especially at low temperature [see the
 dashed curve in \Fig{Fig4}(b)].

\begin{figure}
 \begin{center}
 \includegraphics*[scale=0.5]{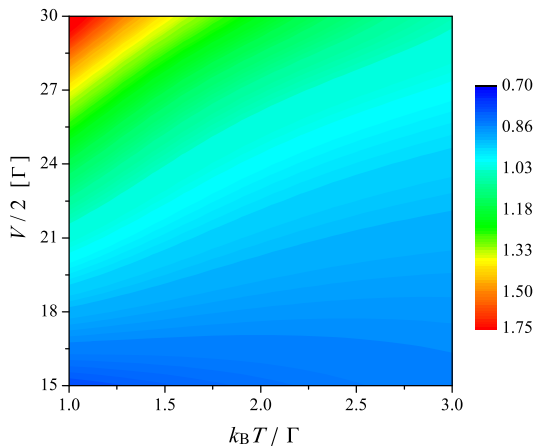}
 \caption{\label{Fig5} Contour plot of the Fano factor versus the
 bias and bare level detuning in the DDCB regime.
 The plotting parameters are, $\GamL=2\GamR$ and $\epl=0$.
 The parameters for the phonon bath are the same as those in \Fig{Fig3}.}
 \end{center}
 \end{figure}

 With the knowledge of energy renormalizations $\tilde{\epl}$
 and $\tilde{\Omg}$, we are now in a position to consider their
 combined effect.
 The Fano factor is plotted as a function of bias voltage
 and temperature in \Fig{Fig5}.
 In the lower half of the bias regime, say $V/2<20\Gam$, the
 level detuning renormalization $\tilde{\epl}$ is not strong
 [see \Fig{Fig2}(a)]. The Fano factor thus exhibits
 sub--Poissonian statistics.
 Increase of $\tilde{\epl}$ due to rising bias voltage results in
 noise enhancement, and consequently super--Poissonian noises are
 observed in the upper half bias regime.
 Noise is further increased at low temperatures, where the interdot
 coupling renormalization $\tilde{\Omg}$ effectively suppresses
 tunneling between the two dots and electrons tend to be
 transferred in bunches more readily.

 \section{\label{thsec5}Conclusion}
 In summary, we have investigated the full counting statistics of
 electron transport through a side--connected double quantum dot
 system, and paid particular attention to the intriguing features
 arising from the renormalized internal dynamics.
 It is found that, in comparison with the average current which
 is not sensitive at all to the detuning renormalization, the Fano
 factor and normalized skewness are both enhanced to a
 super-Poissonian value in a unique way.
 Phonon emission and absorption take place when the double dots
 are further coupled to an external heat bath, and give rise to
 an intriguing interdot coupling renormalization.
 Due to negative contribution, it effectively localizes electron
 in QD2, which eventually leads to strong enhancement of the
 noise spectra.
 The asymmetry of phonon emission and absorption processes result
 in an asymmetric Fano factor, and a change of line shape in the
 skewness.
 Our investigations demonstrate unambiguously the importance
 of the internal energy renormalizations, which thereby should
 be properly accounted for in transport through various nanoscale
 devices.

 \begin{acknowledgments}
 Support from the National Natural Science Foundation of China
 (10904128), Zhejiang Provincial Natural Science Foundation
 (Y6110467), and UGC (AOE/P-04/08-2) and RGC
 (604709) of Hong Kong SAR Government is gratefully acknowledged.
 \end{acknowledgments}


\begin{thebibliography}{10}
\expandafter\ifx\csname url\endcsname\relax
  \def\url#1{\texttt{#1}}\fi
\expandafter\ifx\csname urlprefix\endcsname\relax\def\urlprefix{URL }\fi
\expandafter\ifx\csname href\endcsname\relax
  \def\href#1#2{#2} \def\path#1{#1}\fi

\bibitem{Dat95}
S.~Datta, Electronic Transport in Mesoscopic Systems, Oxford University Press,
  New York, 1995.

\bibitem{Lu03422}
W.~Lu, Z.~Ji, L.~Pfeiffer, K.~W. West, A.~J. Rimberg, Real-time detection of
  electron tunnelling in a quantum dot, Nature 423 (2003) 422.

\bibitem{Fev071169}
G.~F\`{e}ve, A.~Mah\'{e}, J.-M. Berroir, T.~Kontos, B.~Pla\c{c}ais, D.~C.
  Glattli, A.~Cavanna, B.~Etienne, Y.~Jin, An on--demand coherent
  single--electron source, Science 316 (2007) 1169.

\bibitem{Ono021313}
K.~Ono, D.~G. Austing, Y.~Tokura, S.~Tarucha, Current rectification by pauli
  exclusion in a weakly coupled double quantum dot system, Science 297 (2002)
  1313--1317.

\bibitem{Hay03226804}
T.~Hayashi, H.~D.~C. T.~Fujisawa, Y.~Hirayam, Coherent manipulation of
  electronic states in a double quantum dot, Phys. Rev. Lett. 91 (2003) 226804.

\bibitem{Pet052180}
J.~R. Petta, A.~C.Johnson, J.~M.Taylor, E.~A. Laird, A.~Yacoby, M.~D. Lukin,
  C.~M. Marcus, M.~P. Hanson, A.~C. Gossard, Coherent manipulation of coupled
  electron spins in semiconductor quantum dots, Science 309 (2005) 2180.

\bibitem{Kop06766}
F.~H.~L. Koppens, C.~Buizert, K.~J. Tielrooij, I.~T. Vink, K.~C. Nowack,
  T.~Meunier, L.~P. Kouwenhoven, L.~M.~K. Vandersypen, Driven coherent
  oscillations of a single electron spin in a quantum dot, Nature 442 (2006)
  766--771.

\bibitem{Fuj061634}
T.~Fujisawa, T.~Hayashi, R.~Tomita, Y.~Hirayama, "s", Science 312 (2006) 1634.

\bibitem{Gus082547}
S.~Gustavsson, R.~Leturcq, M.~Studer, T.~Ihn, K.~Ensslin, D.~C. Driscoll, A.~C.
  Gossard, Time-resolved detection of single-electron interference, Nano Lett.
  8 (2008) 2547--2550.

\bibitem{Leg871}
A.~J. Leggett, S.~Chakravarty, A.~T. Dorsey, M.~P.~A. Fisher, A.~Garg,
  W.~Zwerger, Dynamics of the dissipative two-state system, Rev. Mod. Phys. 59
  (1987) 1--85, {\bf 67}, 725-726(Erratum) (1995).

\bibitem{Wei08}
U.~Weiss, Quantum Dissipative Systems, World Scientific, Singapore, 2008, 3rd
  ed. Series in Modern Condensed Matter Physics, Vol.\ 13.

\bibitem{Yan05187}
Y.~J. Yan, R.~X. Xu, Quantum mechanics of dissipative systems, Annu. Rev. Phys.
  Chem. 56 (2005) 187--219.

\bibitem{Kon03166602}
J.~{K\"onig}, J.~Martinek, Interaction-driven spin precession in quantum-dot
  spin valves, Phys. Rev. Lett. 90 (2003) 166602.

\bibitem{Bra06075328}
M.~Braun, J.~K{\"{o}nig}, J.~Martinek, Frequency-dependent current noise
  through quantum-dot spin valves, Phys. Rev. B 74 (2006) 075328.

\bibitem{Wun05205319}
B.~Wunsch, M.~Braun, J.~{K\"{o}nig}, D.~Pfannkuche, Probing level
  renormalization by sequential transport through double quantum dots, Phys.
  Rev. B 72 (2005) 205319.

\bibitem{Sot10245319}
B.~Sothmann, J.~{K\"onig}, Transport through quantum-dot spin valves containing
  magnetic impurities, Phys. Rev. B 82 (2010) 245319.

\bibitem{Mar03195305}
F.~Marquardt, C.~Bruder, Dephasing in sequential tunneling through a double-dot
  interferometer, Phys. Rev. B 68 (2003) 195305.

\bibitem{Bru96114}
C.~Bruder, R.~Fazio, H.~Schoeller, Aharonov-bohm oscillations and resonant
  tunneling in strongly correlated quantum dots, Phys. Rev. Lett. 76 (1996)
  114--117.

\bibitem{Fra04085301}
J.~Fransson, O.~Eriksson, Current--voltage asymmetries and negative
  differential conductance due to strong electron correlations in double
  quantum dots, Phys. Rev. B 70 (2004) 085301.

\bibitem{Fra04L85}
J.~Fransson, O.~Eriksson, Asymmetric negative differential conductance in
  double quantum dots, J. Phys.: Condens. Matter 16 (2004) L85--L91.

\bibitem{Luo09385801}
J.~Y. Luo, H.~J. Jiao, F.~Li, X.-Q. Li, Y.~J. Yan, Reduced dynamics with
  renormalization in solid-state charge qubit measurement, J. Phys.: Cond.
  Matt. 21 (2009) 385801.

\bibitem{Luo104904}
J.~Y. Luo, H.~J. Jiao, J.~Z. Wang, Y.~Shen, X.-L. He, Renormalized dynamics in
  charge qubit measurements by a single electron transistor, Phys. Lett. A 374
  (2010) 4904--4908.

\bibitem{Lev964845}
L.~S. Levitov, H.~W. Lee, G.~B. Lesovik, Electron counting statistics and
  coherent states of electron current, J. Math. Phys. 37 (1996) 4845.

\bibitem{Bag03085316}
D.~A. Bagrets, Y.~V. Nazarov, Full counting statistics of charge transfer in
  coulomb blockade systems, Phys. Rev. B 67 (2003) 085316.

\bibitem{Byl05361}
J.~Bylander, T.~Duty, P.~Delsing, Current measurement by real-time counting of
  single electrons, Nature 434 (2005) 361--364.

\bibitem{Sch042005}
R.~Schleser, E.~Ruh, T.~Ihn, K.~Ensslin, D.~C. Driscoll, A.~C. Gossard,
  Time-resolved detection of individual electrons in a quantum dot, Appl. Phys.
  Lett. 85 (2004) 2005--2007.

\bibitem{Van044394}
L.~M.~K. Vandersypen, J.~M. Elzerman, R.~N. Schouten, L.~H. {Willems van
  Beveren}, R.~Hanson, L.~P. Kouwenhoven, Real-time detection of
  single-electron tunneling using a quantum point contact, Appl. Phys. Lett. 85
  (2004) 4394--4396.

\bibitem{Gus06076605}
S.~Gustavsson, R.~Leturcq, B.~Simovic, R.~Schleser, T.~Ihn, P.~Studerus,
  K.~Ensslin, D.~C. Driscoll, A.~C. Gossard, Counting statistics of single
  electron transport in a quantum dot, Phys. Rev. Lett. 96 (2006) 076605.

\bibitem{Gus09191}
S.~Gustavsson, R.~Leturcq, M.~Studer, I.~Shorubalko, T.~Ihn, K.~Ensslin,
  D.~Driscoll, A.~Gossard, Electron counting in quantum dots, Surf. Sci. Rep.
  (2009) 64 (2009) 191--232.

\bibitem{Gru073376}
A.~Gruneis, M.~J. Esplandiu, D.~Garcia-Sanchez, A.~Bachtold, Detecting
  individual electrons using a carbon nanotube field--effect transistor, Nano
  Lett. 7 (2007) 3766--3769.

\bibitem{Thi05045341}
A.~Thielmann, M.~H. Hettler, J.~{K\"{o}nig}, G.~{Sch\"{o}n}, Super-{Poissonian}
  noise, negative differential conductance, and relaxation effects in transport
  through molecules, quantum dots, and nanotubes, Phys. Rev. B 71 (2005)
  045341.

\bibitem{Wel081137}
S.~Welack, J.~B. Maddox, M.~Esposito, U.~Harbola, S.~Mukamel, Single-electron
  counting spectroscopy: Simulation study of porphyrin in a molecular junction,
  Nano Lett. 8 (2008) 1137.

\bibitem{Wie031}
W.~G. van~der Wiel, S.~D. Franceschi, J.~M. Elzerman, T.~Fujisawa, S.~Tarucha,
  L.~P. Kouwenhoven, Electron transport through double quantum dots, Rev. Mod.
  Phys. 75 (2003) 1.

\bibitem{Nau02161303}
A.~Nauen, I.~Hapke-Wurst, F.~Hohls, U.~Zeitler, R.~J. Haug, K.~Pierz, Shot
  noise in self-assembled inas quantum dots, Phys. Rev. B 66 (2002) 161303.

\bibitem{Sas09121926}
S.~Sasaki, H.~Tamura, T.~Akazaki, T.~Fujisawa, Fano-kondo interplay in a
  side-coupled double quantum dot, LANL e-print arXiv:0912.1926.

\bibitem{Li05205304}
X.-Q. Li, J.~Y. Luo, Y.~G. Yang, P.~Cui, Y.~J. Yan, Quantum master-equation
  approach to quantum transport through mesoscopic systems, Phys. Rev. B 71
  (2005) 205304.

\bibitem{Luo07085325}
J.~Y. Luo, X.-Q. Li, Y.~J. Yan, Calculation of the current noise spectrum in
  mesoscopic transport: A quantum master equation approach, Phys. Rev. B 76
  (2007) 085325.

\bibitem{Luo08345215}
J.~Y. Luo, X.-Q. Li, Y.~J. Yan, Spin-dependent current noises in transport
  through coupled quantum dots, J. Phys.: Cond. Matt. 20 (2008) 345215.

\bibitem{Wan07125416}
S.-K. Wang, H.~Jiao, F.~Li, X.-Q. Li, Y.~J. Yan, Full counting statistics of
  transport through two-channel coulomb blockade systems, Phys. Rev. B 76
  (2007) 125416.

\bibitem{Gro06125315}
C.~W. Groth, B.~Michaelis, C.~W.~J. Beenakker, Counting statistics of coherent
  population trapping in quantum dots, Phys. Rev. B 74 (2006) 125315.

\bibitem{Fli05475}
C.~Flindt, T.~Novotny, A.-P. Jauho, Full counting statistics of
  nano-electromechanical systems, Europhys. Lett. 69 (2005) 475.

\bibitem{Kie06033312}
G.~{Kie{\ss}lich}, P.~Samuelsson, A.~Wacker, E.~{Sch\"{o}ll}, Counting
  statistics and decoherence in coupled quantum dots, Phys. Rev. B 73 (2006)
  033312.

\bibitem{Gur9615932}
S.~A. Gurvitz, Y.~S. Prager, Microscopic derivation of rate equations for
  quantum transport, Phys. Rev. B 53 (1996) 15932--15943.

\bibitem{Sto961050}
T.~H. Stoof, Y.~V. Nazarov, Time-dependent resonant tunneling via two discrete
  states, Phys. Rev. B 53 (1996) 1050--1053.

\bibitem{Bra05315}
T.~Brandes, Coherent and collective quantum optical effects in mesoscopic
  systems, Phys. Rep. 408 (2005) 315--474.

\bibitem{Kie07206602}
G.~{Kie{\ss}ich}, E.~Sch{\"{o}}ll, T.~Brandes, F.~Hohls, R.~J. Haug, Noise
  enhancement due to quantum coherence in coupled quantum dots, Phys. Rev.
  Lett. 99 (2007) 206602.

\bibitem{Agu04206601}
R.~Aguado, T.~Brandes, Shot noise spectrum of open dissipative quantum
  two-level systems, Phys. Rev. Lett. 92 (2004) 206601.

\bibitem{Luo10083720}
J.~Y. Luo, S.-K. Wang, X.-L. He, X.-Q. Li, Y.~J. Yan, Real-time counting of
  single electron tunneling through a t-shaped double quantum dot system, J.
  Appl. Phys. 108 (2010) 083720.

\bibitem{Gra80}
I.~S. Gradshteyn, I.~M. Ryzhik, Table of Integrals, Series, and Products,
  Academic Press, New York, 1980, 4th ed.

\bibitem{Gur9715215}
S.~A. Gurvitz, Measurements with a noninvasive detector and dephasing
  mechanism, Phys. Rev. B 56 (1997) 15215--23.

\bibitem{Gur986602}
S.~A. Gurvitz, Rate equations for quantum transport in multidot systems, Phys.
  Rev. B 57 (1998) 6602--11.

\bibitem{Haz01165313}
B.~L. Hazelzet, M.~R. Wegewijs, T.~H. Stoof, Y.~V. Nazarov, Coherent and
  incoherent pumping of electrons in double quantum dots, Phys. Rev. B 63
  (2001) 165313.

\bibitem{Bla001}
Y.~M. Blanter, M.~{B\"{u}ttiker}, Phys. Rep. 336 (2000) 1.

\bibitem{Suk01125315}
E.~V. Sukhorukov, G.~Burkard, D.~Loss, Noise of a quantum dot system in the
  cotunneling regime, Phys. Rev. B 63 (2001) 125315.

\bibitem{Bra091048}
A.~Braggio, C.~Flindt, T.~{Novotn\'{y}}, The influence of charge detection on
  counting statistics, J. Stat. Mech. 1 (2009) 1048.

\bibitem{Urb09165319}
D.~Urban, J.~{K\"{o}nig}, Tunable dynamical channel blockade in double-dot
  {Aharonov-Bohm} interferometers, Phys. Rev. B 79 (2009) 165319.

\end{thebibliography}

\end{document}